\documentclass[preprintnumbers,aps,prb,twocolumn,superscriptaddress,groupedaddress]{revtex4}

\usepackage{graphicx}
\usepackage{dcolumn} 
\usepackage{amsmath}
\usepackage{exscale}
\usepackage{bm}      
\usepackage{color}
\usepackage{epsfig,psfrag,subfigure,subeqnarray}
\usepackage{bbm}
\usepackage{amsbsy}
\usepackage{amssymb}
\usepackage{enumitem}
\usepackage{mathrsfs}

\def\lan{\langle}
\def\ran{\rangle}
\def\va{\varepsilon}

\def\vk{{\bf k}}
\def\vK{{\bf K}}

\def\vr{{\bf r}}

\def\vp{{\bf p}}

\def\v0{{\bf 0}}
\newcommand{\bd}{\begin{equation}}
\newcommand{\ed}{\end{equation}}
\newcommand{\be}{\begin{equation}}
\newcommand{\ee}{\end{equation}}
\newcommand{\bt}{\begin{split}}
\newcommand{\et}{\end{split}}

\newcommand{\bn}{\begin{align}}
\newcommand{\en}{\end{align}}

\newcommand{\bea}{\begin{eqnarray}}
\newcommand{\eea}{\end{eqnarray}}
\newcommand{\ba}{\begin{array}}
\newcommand{\ea}{\end{array}}
\newcommand{\nn}{\nonumber}

\begin{document}

\title{Understanding Semiconductor Valence Mass}

\author{Monique Combescot}
\affiliation{Sorbonne Universit\'e, CNRS, Institut des NanoSciences de Paris, 75005-Paris, France}
\author{Shiue-Yuan Shiau}
\affiliation{Physics Division, National Center for Theoretical Sciences, Hsinchu, 30013, Taiwan}

\begin{abstract}
The Bloch theorem mathematically proves that in a periodic crystal, electrons can acquire a negative mass.
 The present work aims to provide a physical understanding for why this is so. We successively analyze the consequences of the 3-fold orbital valence state coupling to (i) a non-degenerate orbital level in the conduction band, (ii) a 3-fold orbital level in the conduction band, and (iii)  spin states through spin-orbit interaction. We show that it is not at all trivial for valence electrons to acquire a negative mass for \textit{whatever} their momentum  with respect to the crystal axes: it is necessary to not only have a coupling to a degenerate orbital conduction level, but also a symmetry breaking of the 3-fold valence subspace by the spin quantization axis, as induced by spin-orbit interaction. Due to the relativistic origin of this interaction, the  
existence of negative valence masses thus constitutes an unexpected  
signature of quantum relativity.
\end{abstract}
\date{\today}

\maketitle

\section{Introduction\label{sec1}}

In semiconductors, some electrons have a negative mass. This fact is the prime reason for these materials to have shaped today's technology\cite{Esaki1970,Capasso1987}. As a physicist, it is worth understanding why and how this is so, independently from the mathematical derivations based on Bloch theorem\cite{Bloch,Kittelbook,Merminbook} or group theory\cite{Falicov,Bir,Ivchenko}, and also independently from the numerous methods that calculate semiconductor band structures, like the tight-binding approach\cite{Vogl,Sarma}, the Luttinger's approach\cite{Luttinger} and the numerical density functional method\cite{Kohn1965,Argaman}.


A direct consequence of electrons having a negative effective mass is that when such an electron is excited, it leaves a hole, that is, an electron absence in the corresponding state\cite{Kittelbook,Merminbook}. This hole essentially behaves as a quantum particle having a positive mass. As a result, semiconductors  host two types of fermionic particles, the conduction electrons and the valence holes. Having opposite charges, they can bind into a bosonic particle, the exciton, which is similar to a hydrogen atom, but with a much larger size due to the electron and hole effective masses, one order of magnitude smaller than the free electron mass, and to the semiconductor dielectric constant of the order of 10. These basics largely explain that in addition to their tremendous technological interest\cite{Capasso1987,Gmachl,Sang}, semiconductors have provided an ideal playground for a large number of  exciting many-body effects of fundamental physics\cite{Bastardbook,Haugbook,Cardona,Monicbook}.

The Bloch theorem mathematically proves that the energies of electrons subjected to a periodic ion lattice potential form bands, with minima and maxima. Close to these extrema, the band curvatures are  positive or negative, enforcing the resulting electron effective mass to also be positive or negative, as commonly demonstrated in textbooks through the simple Kronig-Penney model\cite{Kronig}. Behind this beautiful but abstruse mathematics, there is a drastic  change in physics: the electron that suffers a periodic potential is a free electron with  a positive mass $m_0$. Due to interaction with the periodic ion lattice, this electron ends by behaving like a free electron but with a much lighter mass, that can even become negative. What are the forces that  drive this pathological sign change in the electron mass? Understanding its physics  will help tailoring the particle effective mass of other systems in search for new technology\cite{Silveirinha2012,ChangPRL,Khamehchi}.

The usual approach to the effective masses of semiconductor electrons makes use of group theory. Indeed, the degenerate orbital states of the upper valence band suffer a relativistic spin-orbit interaction\cite{Thomas,MonicPRB2019} that mixes orbital and spin degrees of freedom. A standard but simple-minded way to derive the spin-orbit eigenstates for semiconductors follows the procedure developed for atoms\cite{Cardona}. It is based on the total angular momentum $\textbf{J}=\textbf{L}+\textbf{S}$ of the electron. However, because the concept of orbital momentum $\textbf{L}$ only has a meaning for electron states that have a spherical symmetry\cite{cohenbook}, like atomic states, such a derivation cannot be used for crystals, in spite of the validity of the obtained results. Up to very recently\cite{MonicPRB2019}, the procedure based on group theory has been the only correct way to derive the spin-orbit eigenstates in a periodic crystal. Through the double groups that mix orbital and spin symmetries, this approach leads to different dispersion relations for valence electrons in the various spin-orbit states.

For sure, the group theory formalism is extremely powerful. Yet, through mixing the orbital and spin subspaces within the double groups, as done at the very first line, it lacks a physical transparency. In addition,  most semiconductor physicists who do not master group theory find it difficult to follow, and equally hard to accept that learning such a general but heavy formalism is necessary to solve problems dealing with a 3-fold orbital degeneracy only.  This is why, in a first work\cite{MonicPRB2019}, we reconsidered the energy splitting of the 3-fold orbital states at the $\Gamma$ point, \textit{i.e.}, for $\vk=\textbf{0}$ electron momentum, induced by the spin-orbit interaction. We pinned down the physical origin of the change from the ``natural" orbital states $\mu=(x,y,z)$ labeled along the crystal axes, to their linear combinations $\eta=(1,0,-1)$ similar to the $\ell=1$ atomic states.\

 The present work follows the same spirit: by only using conceptually simple arguments, we unravel the physics that drives the positive mass of a free electron toward a negative  value when this electron is put in a periodic ion lattice. 
The sign change in the electron mass, from positive to negative, fundamentally relies on couplings to states that are \textit{opposite} in parity and \textit{higher} in energy than the upper valence states at hand. Yet, this is far from enough. To end with a negative effective mass \textit{whatever} the electron momentum direction $\vk$ with respect to the crystal axes, it is necessary to break the orbital symmetry of the degenerate valence states along these axes. This is done in two ways:

(i) By mixing  the orbital states of the degenerate valence level, through its coupling to the orbital states of a \textit{degenerate} level in the conduction band. By contrast, its coupling to a non-degenerate conduction level, like the  lowest conduction band, plays no role in getting a negative mass.

(ii) By introducing the electron spin. Through the spin-orbit interaction, the spin quantization axis breaks the orbital degeneracy $\mu=(x,y,z)$ of electrons in a cubic crystal. The resulting orbital eigenstates, labeled as
 $\eta=\pm 1$, are linear combinations of $(x,y)$ states, while the $z$ state along the spin quantization axis stays unaffected: it just corresponds to the spin-orbit eigenstate labeled as $\eta=0$. 

These two ways of mixing are  necessary for valence electrons to end with a negative effective mass whatever the $\vk$ momentum direction. It is then clear that their interplay renders the physics of this negative effective mass quite complicated, and definitely beyond the simple interpretation based on a Kronig-Penney model\cite{Kronig}, even if the Bloch theorem mathematically draws the correct conclusion within a few lines. Indeed, the common textbook explanation for negative effective mass involves Bragg reflection: In one-dimensional spinless Kronig-Penney model with a lattice spacing $a$, negative effective mass occurs when the momentum $k$ approaches the band boundaries $n\pi/a$ from below, with $n=\pm1,\pm2,\cdots$.  When $k$ is slightly less than $n\pi/a$, the scattering processes from each lattice potential interfere in such a way that the particle wave begins to be largely reflected, that is, propagate backward even as we increase its momentum $k$ by applying a force; the particle transfers more momentum to the lattice than it receives from the applied force\cite{Kittelbook}. While this simple physical argument supports the negative effective mass as a general feature in all periodic crystals, it is obvious that the physics of negative effective mass in real materials is more involved.

In order to catch the above physics in the simplest way, we here use a $\vk\cdot\vp$ approach\cite{Voonbook,Cavassilas,Fishman}. Its big advantage is to possibly analyze, step by step, the various couplings experienced by the valence electrons, to ultimately catch the key ones.

This paper is organized as follows: 

Section \ref{sec2} provides general arguments on the physical origin of a negative valence mass and the possible dependence of the electron energy on the momentum $\vk$, making use of the fact that in a cubic crystal, the $(x,y,z)$ axes play the same role. 

Section \ref{sec3} recalls the $\vk\cdot\vp$ procedure to obtain the electron energy up to second order in momentum $\vk$. We give the possible forms for the degenerate and non-degenerate orbital wave functions of a cubic crystal, which fundamentally are even in the case of the valence band and odd in the case of the conduction band. We also give the form of the eigenstates resulting from spin-orbit interaction. Finally, for later use, we relate operators written in the $\mu=(x,y,z)$ basis and the $\eta=(1,0,-1)$ basis.
 
  In Section~\ref{sec4}, we derive the energy of a valence electron in a 3-fold orbital level, by considering its coupling to an orbital level in the conduction band which is either non-degenerate or 3-fold, or its coupling to both types of levels.

   In Section~\ref{sec5}, we perform similar calculations, focusing on the 4-fold spin-orbit eigenstates, labeled as $\tilde{\jmath}_z=(\pm3/2,\pm 1/2)$, which are the physically relevant states in the presence of spin-orbit interaction.

We then conclude.  

\section{General arguments\label{sec2}}
We consider a direct-gap semiconductor like GaAs, with band extrema occurring for $\vk=\textbf{0}$, called  $\Gamma$ point. We want to determine the dispersion relation close to the maximum of the valence band that originates from  3-fold orbital states. Due to symmetry between the cubic crystal axes $(x,y,z)$, we expect the dispersion relation, that is, the energy difference $\va_{v,\vk}-\va_{v,\textbf{0}}$, to depend on the components of the momentum $\vk$ through $(k_x^2,k_y^2,k_z^2)$ taken in a cyclic way. Possible combinations at lowest order in $k$ are $k_x^2+k_y^2+k_z^2=k^2$, 
\be
k_x^2k_y^2+k_y^2k_z^2+k_z^2k_x^2=S_\vk\,, \,\,\,\,\,\,\,
\quad  k_x^2k_y^2k_z^2=P_\vk\,.\label{k.p_1}
\ee   
Terms like $S_\vk$ or $P_\vk$ in the dispersion relation cause the energy to be non-spherical---known as ``warping.'' Handling such a non-spherical energy in problems dealing with Coulomb interaction is fundamentally impossible. This is why the warping is commonly averaged out to produce a spherical mass. We will however see that the existence of a warping in the valence band is fundamentally linked to the existence of a negative curvature: it is a crucial feature of the valence effective mass problem.

\section{Required background\label{sec3}}

The derivation we propose for the valence electron effective mass does not require the knowledge of group theory. It only uses very basic solid state physics. Nevertheless, to settle the notations properly, we have chosen to briefly recall what we are going to use.

\subsection{The $\vk\cdot\vp$ formalism}
The $\vk\cdot\vp$ formalism\cite{Voonbook} is a  simple but  powerful tool to understand semiconductor band structures. It starts with the Hamiltonian of a free electron with positive mass $m_0$, in a potential $V(\vr)$ having the lattice periodicity. The Bloch states $|n,\vk\ran$, eigenstates of this Hamiltonian 
\be
0= \left(\frac{\hat{\vp}^2}{2m_0}+V(\vr)-\va_{n,\vk}\right)|n,\vk\ran\,,
\label{k.p_2}
\ee
are characterized by a band index $n$ and a momentum $\vk$.
 When $\vk$ is small, the $\va_{n,\vk}$ energy can be obtained from the knowledge of all Bloch states for $\textbf{k}=\textbf{0}$. Indeed, by writing the Bloch state wave function in a sample volume $L^3$ as
\be
\lan \vr|n,\vk\ran=\frac{e^{i\vk\cdot\vr}}{L^{3/2}}\lan \vr|u_{n,\vk}\ran\label{k.p_3}
\ee
in Eq.~(\ref{k.p_2}), we find that $|u_{n,\vk}\ran$ fulfills
\be
0= (\hat{h}_\vk-\va_{n,\vk})|u_{n,\vk}\ran\,,\label{k.p_4}
\ee
with $\hat{h}_\vk=\hat{h}_\v0+\hbar^2\vk^2/2m_0+\hat{w}_\vk$. For $\vk$ small, the operator $\hat{w}_\vk$, defined as
\be
\hat{w}_\vk=\frac{\hbar}{m_0}\vk\cdot\hat{\vp}\,,\label{k.p_5}
\ee
can be treated as a perturbation.

\noindent (i) For $\va_{n_0,\v0}$ non-degenerate, the $\va_{n_0,\vk}$ energy follows from
\bea
\tilde{\va}_{n_0,\vk}&=&\va_{n_0,\vk}-\va_{n_0,\v0}-\frac{\hbar^2\vk^2}{2m_0}\nn\\
&\simeq&\lan u_{n_0,\v0}|\hat{w}_\vk P_\perp \frac{1}{\va_{n_0,\v0}-\hat{h}_{\v0}} P_\perp \hat{w}_\vk | u_{n_0,\v0}\ran\,,\label{k.p_6}
\eea
where $P_\perp$  is the projector over the subspace  orthogonal to $|u_{n_0,\v0}\ran$, namely $P_\perp=\sum_{n\not=n_0} |u_{n,\v0}\ran \lan u_{n,\v0}|$. Note that the first-order term $\lan u_{n_0,\v0}|\hat{w}_\vk| u_{n_0,\v0}\ran$ disappears due to parity.

\noindent (ii) When $\va_{n_0,\v0}$ is degenerate, that is, when $\va_{n_0,\v0}$ is the energy of $|u^{(r)}_{n_0,\v0}\ran$ states with $r=(1,2,\ldots, N_0)$, the $\hat{w}_\vk$ operator shifts and splits these $N_0$ states. The $\va_{n_0,\vk}$ eigenvalues then follow from the cancellation of the determinant of a $N_0\times N_0$ matrix $\hat{\mathcal{D}}$  defined as
\bea
\mathcal{D}_{r',r}&=& -\tilde{\va}_{n_0,\vk}\delta_{r',r}+\mathcal{W}_{r',r}\,,\label{k.p_7}\\
\mathcal{W}_{r',r}&=&\lan u^{(r')}_{n_0,\v0}|\hat{w}_\vk P_\perp \frac{1}{\va_{n_0,\v0}-\hat{h}_{\v0}} P_\perp \hat{w}_\vk | u^{(r)}_{n_0,\v0}\ran\,,\label{k.p_8}
\eea
 where $P_\perp$ now is the projector over the subspace orthogonal to the $|u^{(r)}_{n_0,\v0}\ran$ subspace. In practice, calculations are made with a restriction of this orthogonal subspace to states whose  energy is close to $\va_{n_0,\v0}$, due to the $(\va_{n_0,\v0}-\hat{h}_{\v0})$ denominator in the above equation.

Without going into detailed calculation, some useful observations can be drawn from the above equations.
 
\noindent (i) The states coupled to $|u^{(r)}_{n_0,\v0}\ran$ through $\hat{w}_\vk$, that enter the $P_\perp$ operator, lead to  negative $\mathcal{W}_{r',r}$'s when they are higher in energy than  $\va_{n_0,\v0}$. They produce a negative $\tilde{\va}_{n_0,\vk}$ value, that makes the resulting $\va_{n_0,\vk}$ curvature, \textit{i.e.}, the inverse effective mass, smaller than $1/2m_0$, and possibly negative when the coupling is large. 

\noindent (ii) When the $\hat{\mathcal{W}}$ determinant is equal to zero, the equation $\text{det} \hat{\mathcal{D}}=0$ must contain a solution  $\tilde{\va}_{n_0,\vk}=0$, that is, $\va_{n_0,\vk}=\va_{n_0,\v0}+\hbar^2\vk^2/2m_0$, which  corresponds to an effective mass equal to the positive free electron mass $m_0$. As a result, when $\text{det} \hat{\mathcal{W}}=0$, we immediately deduce that the amount of states included in the truncated $P_\perp$ operator is not enough to produce a negative effective mass, as required for valence electrons.

\begin{figure}[t]
\begin{center}
\includegraphics[trim=6cm 5.5cm 6cm 4cm,clip,width=2.2in]{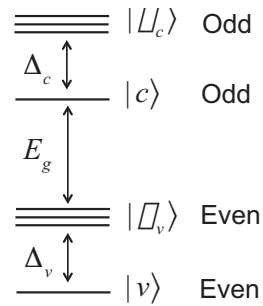}
\caption{The physically relevant orbital states are the ones close to the band gap $E_g$. In zinc-blend semiconductors like GaAs, these are the non-degenerate and 3-fold degenerate conduction states $|c\ran$ and $|\mu_c\ran$ separated by an energy $\Delta_c$, and the non-degenerate and 3-fold degenerate valence states $|v\ran$ and $|\mu_v\ran$ separated by an energy $\Delta_v$. Conduction states are fundamentally odd in parity, while valence states are even.}
\label{fig:1}
\end{center}
\end{figure}

\subsection{Valence and conduction states with crystal periodicity\label{sec1.2}}

\noindent $\bullet$ The physically relevant states of the valence and conduction bands are the ones close to the band gap. They are shown in Fig.~\ref{fig:1}. Their orbital parts are either non-degenerate as in the case of the $|v\ran$ and $|c\ran$ states, or 3-fold degenerate as in the case of $|\mu_v\ran$ and $|\mu_c\ran$ states, with $\mu=(x,y,z)$ along the crystal axis\cite{Cavassilas}. Having the lattice periodicity, we can expand their wave functions in terms of the reciprocal lattice vectors $\vK$.

Conduction orbital states have a parity which is fundamentally odd. Since the cubic axes play the same role, we can write them\cite{MonicPRB2019} as
\bea
&&-\lan -\vr|c\ran=\lan \vr|c\ran=\sum_\vK \frac{e^{i\vK\cdot\vr}}{L^{3/2}}K_xK_yK_z\, F_{K,c}\,,\label{k.p_9}\\
&&-\lan -\vr|\mu_c\ran=\lan \vr|\mu_c\ran=\sum_\vK \frac{e^{i\vK\cdot\vr}}{L^{3/2}}K_\mu \,G_{K,c}\,.\label{k.p_10}
\eea
Similarly, the valence orbital states, which are fundamentally even, can be written as
\bea
&&\lan -\vr|v\ran=\lan \vr|v\ran=\sum_\vK \frac{e^{i\vK\cdot\vr}}{L^{3/2}} \,\,F_{K,v}\,,\label{k.p_11}\\
&&\lan -\vr|\mu_v\ran=\lan \vr|\mu_v\ran=\sum_\vK \frac{e^{i\vK\cdot\vr}}{L^{3/2}}\frac{K_xK_yK_z}{K_\mu} \,\,G_{K,v}\,.\label{k.p_12}
\eea
These $F$ and $G$ functions only depend on $K=|\vK|$.

\begin{figure}[t]
\begin{center}
\includegraphics[trim=6cm 5.5cm 4cm 4cm,clip,width=2.6in]{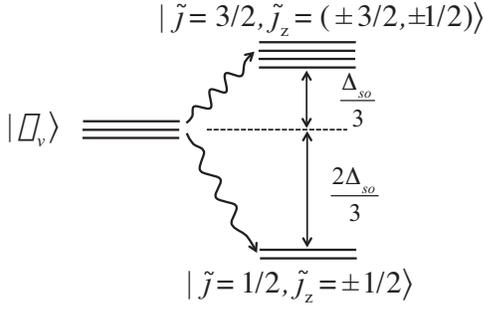}
\caption{The spin-orbit interaction splits the six valence states $|\mu_v\ran\otimes |\pm1/2\ran$ into four and two states, separated by $\Delta_{so}$. These states are given in Eqs.~(\ref{k.p_13},\ref{k.p_14},\ref{k.p_15}).}
\label{fig:2}
\end{center}
\end{figure}

\noindent $\bullet$ The cyclic symmetry of these orbital states is broken by the spin quantization axis when spin-orbit interaction is introduced. The six states $|\mu_v\ran\otimes |\pm1/2\ran$ then split into four and two states (see Fig.~\ref{fig:2}) in a way similar to their atomic counterpart with quantum number $j$. This is why we label them as $\tilde{\jmath}$, not $j$, in order to keep the reader reminded that orbital angular momentum has no meaning for electrons in a periodic crystal.

The  upper 4-fold states read as 
\bea
\label{k.p_13}
|\tilde{\jmath}=3/2,\tilde{\jmath}_z=\pm3/2\ran=|\pm1_v\ran\otimes |\pm1/2\ran\,,\,\,\,\,\,\,\,\,\,\,\,\,\,\,\,\,\,\,\,\,\,\,\,\,\,\,\,\,\,\,
\\
|\tilde{\jmath}=3/2,\tilde{\jmath}_z=\pm1/2\ran=\,\,\,\,\,\,\,\,\,\,\,\,\,\,\,\,\,\,\,\,\,\,\,\,\,\,\,\,\,\,\,\,\,\,\,\,\,\,\,\,\,\,\,\,\,\,\,\,\,\,\,\,\,\,\,\,\,\,\,\,\,\,\,\,\,\,\,\,\,\,\,\,\,\,\,\,\,\,\,
\label{k.p_14}
\\
\nn
\frac{1}{\sqrt{3}}\left(|\pm1_v\ran\otimes |\mp1/2\ran+\sqrt{2}|0_v\ran\otimes |\pm1/2\ran\right)\,,\,
\eea
while the lower 2-fold states read as
\bea
|\tilde{\jmath}=1/2,\tilde{\jmath}_z=\pm1/2\ran=\,\,\,\,\,\,\,\,\,\,\,\,\,\,\,\,\,\,\,\,\,\,\,\,\,\,\,\,\,\,\,\,\,\,\,\,\,\,\,\,\,\,\,\,\,\,\,\,\,\,\,\,\,\,\,\,\,\,\,\,\,\,\,\,\,\,
\label{k.p_15}\\
\frac{1}{\sqrt{3}}\left(\sqrt{2}|\pm1_v\ran\otimes |\mp1/2\ran-|0_v\ran\otimes |\pm1/2\ran\right)\,.
\nn
\eea

 For a spin quantization axis taken along $\textbf{z}$, the orbital states $|\eta_v\ran$  with $\eta=(1,0,-1)$ read in terms of the $|\mu_v\ran$ orbital states as 
\be
|\pm1_v\ran=\frac{\mp i|x_v\ran+|y_v\ran }{\sqrt{2}},\,\,\,\,\,\quad |0_v\ran=i|z_v\ran\,,\label{k.p_16}
\ee
following the Landau-Lifshitz choice\cite{Landau} for the arbitrary phase factor.

\subsection{From $(x,y,z)$ to $(1,0,-1)$ orbital states}
The above results show that the relevant orbital indices to handle spin-orbit interaction are not the ones linked to the crystal axes because their cyclic symmetry is broken by the spin quantization axis. Yet, this cyclic symmetry renders the $(x,y,z)$ indices quite convenient for calculations before  introducing the spin. This is why we will start with operator written in the $(x,y,z)$ basis and then turn to its representation in the $(1,0,-1)$ basis. 

The $3\times 3$ matrix that represents the operator $\hat{A}$ in the $\mu=(x,y,z)$ state basis reads as
\be
\hat{A}_\mu=\begin{pmatrix}
A_{x,x} & A_{x,y} & A_{x,z} \\
A_{y,x} & A_{y,y} & A_{y,z}\\
A_{z,x} & A_{z,y} & A_{z,z}
\end{pmatrix}\,,\label{k.p_17}
\ee
with $A_{\mu',\mu}=\lan\mu'|\hat{A}|\mu\ran$, equal to $A^*_{\mu,\mu'}$ when $\hat{A}=\hat{A}^\dagger$. In the problem we here consider, these matrix elements are real; so, $A_{\mu',\mu}=A_{\mu,\mu'}$, as taken in the following.

In the $\eta=(1,0,-1)$ basis, the same operator reads as 
\be
\hat{A}_\eta=\begin{pmatrix}
A_{1,1} & A_{1,0} & A_{1,-1} \\
A_{0,1} & A_{0,0} & A_{0,-1}\\
A_{-1,1} & A_{-1,0} & A_{-1,-1}
\end{pmatrix}\,,\label{k.p_18}
\ee
with $A_{\eta',\eta}=\lan \eta'|\hat{A}|\eta\ran$ still equal to $A^*_{\eta,\eta'}$ because $\hat{A}=\hat{A}^\dagger$. But $A_{\eta',\eta}$ is not necessarily equal to $A_{\eta,\eta'}$ because the matrix elements of $\hat{A}$, real in the $\mu$ basis, are not necessarily real in this $\eta$ basis.

We can relate these two sets of matrix elements by inserting the closure relation $\sum_\mu|\mu\ran\lan \mu|$ on both sides of the operator $\hat{A}$ as
\be
A_{\eta',\eta}=\lan \eta'|\hat{A}|\eta\ran=\lan \eta'|\sum_{\mu'}|\mu'\ran\lan \mu'|\hat{A}\sum_\mu|\mu\ran\lan \mu|\eta\ran\,.\label{k.p_19}
\ee
The link between the $|\eta\ran$ and  $|\mu\ran$ states given in Eq.~(\ref{k.p_16}) leads to 
\bea
A_{1,1}=\frac{i\lan x_v|+\lan y_v|}{\sqrt{2}}\hat{A}\frac{-i|x_v\ran+|y_v\ran}{\sqrt{2}}=\frac{A_{x,x}+A_{y,y}}{2}\,,\label{k.p_20}\\
A_{\pm1,0}=\frac{\pm i\lan x_v|+\lan y_v|}{\sqrt{2}}\hat{A}i|z_v\ran=\frac{\mp A_{x,z}+iA_{y,z}}{\sqrt{2}}\,,\label{k.p_21}
\eea
and so on... So, the $\hat{A}$ operator in the $(1,0,-1)$ basis reads as
\be
\hat{A}_\eta=\begin{pmatrix}
\frac{A_{x,x}+A_{y,y}}{2} & \frac{-A_{x,z}+iA_{y,z}}{\sqrt{2}} & \frac{-A_{x,x}+A_{y,y}+2i A_{x,y}}{2}\\
\cdot & A_{z,z} & \frac{A_{x,z}-iA_{y,z}}{\sqrt{2}}\\
\cdot & \cdot& \frac{A_{x,x}+A_{y,y}}{2}
\end{pmatrix}\,,
\label{k.p_22}
\ee
the other elements being obtained from $A^*_{\eta',\eta}=A_{\eta,\eta'}$.

The determinants of the two matrices $\hat{A}_\mu$ and $\hat{A}_\eta$ are equal because these two matrices represent the same operator $\hat{A}$ in different bases of the same $3\times3$ subspace. Yet, the cyclic symmetry of the $(x,y,z)$ indices renders the calculation of this determinant far simpler for $\hat{A}_\mu$.

\section{Valence electron effective mass in the absence of spin\label{sec4}}

We first forget spin and look for the effect of the $\hat{w}_\vk$ operator on the 3-fold orbital states $|\mu_v\ran$ given in Sec.~\ref{sec3}. Since the $\hat{\vp}=-i\hbar\nabla$ operator in $\hat{w}_\vk$ is odd, it couples the even valence states $|\mu_v\ran$ to odd states only; so, it does not couple $|\mu_v\ran$ to other valence states, neither $|v\ran$ nor $|\mu'_v\ran$ with $\mu'  \neq \mu$, but to the conduction state $|c\ran$ and possibly to the $|\mu_c\ran$ states slightly above in energy (see Fig.~\ref{fig:1}).

\subsection{Coupling to $|c\ran$ only}
\noindent $\bullet$ Using the periodic wave functions given in Sec.~\ref{sec1.2}, we find
\bea
\lan c|\hat{p}_x|\mu_v\ran&=&\int d^3r \sum_{\vK'}\frac{e^{-i\vK'\cdot\vr}}{L^{3/2}}K'_xK'_yK'_z\, F^*_{K',c}\label{k.p_23}\\
&&\times\frac{\hbar}{i}\frac{\partial}{\partial x}\left(\sum_\vK \frac{e^{i\vK\cdot\vr}}{L^{3/2}}\frac{K_xK_yK_z}{K_\mu} \,G_{K,v}\right)\,.\nn
\eea
As $\frac{\partial}{\partial x}e^{i\vK\cdot\vr}=iK_x e^{i\vK\cdot\vr}$, the integral over $\vr$ gives $L^3\delta_{\vK',\vK}$; so, the RHS of the above equation reduces to 
\be
\hbar \sum_\vK \frac{K_x}{K_\mu} K_x^2K_y^2K_z^2\, F^*_{K,c}\,G_{K,v}\,,\label{k.p_24}
\ee
which, for a cubic crystal, is equal to zero except for $\mu=x$. So, 
\be
\lan c|\hat{p}_x|\mu_v\ran=\delta_{x,\mu}\hbar \sum_\vK K_x^2K_y^2K_z^2 \,F^*_{K,c}\,G_{K,v}  \equiv \delta_{x,\mu}\mathcal{P}_{cv}\,.\label{k.p_25}
\ee

\noindent $\bullet$ The matrix elements of the $\hat{w}_\vk$ operator between the 3-fold valence states $|\mu_v\ran$ and the non-degenerate conduction state $|c\ran$ then reduces to
\bea
 \lan c|\hat{w}_\vk|\mu_v\ran&=&\lan c|\frac{\hbar}{m_0}(k_x\hat{p}_x+k_y\hat{p}_y+k_z\hat{p}_z)|\mu_v\ran\nn\\
 &=&\frac{\hbar\mathcal{P}_{cv}}{m_0}k_\mu\,.\label{k.p_26}
\eea

\noindent $\bullet$ When the $|c\ran$ state only is included in the $P_\perp$ projector appearing in Eq.~(\ref{k.p_8}), we get
\bea
\label{k.p_27}
\mathcal{W}_{\mu',\mu}&=&\lan \mu'_v|\hat{w}_\vk\frac{|c\ran\lan c|}{-E_g}\hat{w}_\vk|\mu_v\ran
\\
&=&-\frac{\hbar^2 |\mathcal{P}_{cv}|^2}{m_0^2E_g}k_{\mu'}k_{\mu}
\equiv-\frac{\hbar^2}{2m_0}\gamma_1\lan \mu'_v|\hat{\mathcal{B}}^{(1)}|\mu_v\ran\,,\nn
\eea
where $\gamma_1$ is a dimensionless parameter given by
\be
\gamma_1=\frac{2|\mathcal{P}_{cv}|^2}{m_0E_g}\,.
\ee 
The corresponding $3 \times 3$ matrix 
for $\hat{\mathcal{W}}$ then reads 
$-(\hbar^2 \gamma_1 /2m_0) \hat{\mathcal{B}}^{(1)}$ where $\hat{\mathcal{B}}^{(1)}$ is a 
 symmetric real matrix 
\be
\hat{\mathcal{B}}^{(1)}_\mu=\begin{pmatrix}
k_x^2 & \cdot & \cdot \\
k_xk_y & k_y^2 & \cdot\\
k_zk_x &k_yk_z & k_z^2
\end{pmatrix}\,.\label{k.p_28}
\ee

Since $\text{det}\hat{\mathcal{B}}^{(1)}=0$, as easy to check, we readily see that the equation $\text{det}\hat{\mathcal{D}}=0$, for coupling to $|c\ran$ only,  
has a solution  $\tilde{\va}_{v,\vk}=0$, which corresponds to  an unchanged positive mass $m_0$. More precisely, the solutions of $\text{det}\hat{\mathcal{D}}=0$ read as $\tilde{\va}_{v,\vk}= (\hbar^2/ 2m_0)   \gamma_1 e_1$ with $ e_1$ solution of
\be
0=\left|\begin{matrix}
k_x^2+e_1 &   \cdot  & \cdot  \\
k_xk_y & k_y^2+e_1 &  \cdot  \\
k_zk_x &k_yk_z & k_z^2+e_1
\end{matrix}\right|=e_1^2(e_1+k^2)\,.\label{k.p_29}
\ee
So, the coupling to $|c\ran$ only leads to a partial splitting of the valence orbital states $|\mu_v\ran$, with two unchanged branches still having a positive mass $m_0$ which corresponds to the $e_1=0$ degenerate solution, and one branch  $e_1=-k^2$ which gives
\be
\va_{v,\vk}=\va_{v,{\bf0}}+(1-\gamma_1)\frac{\hbar^2k^2}{2m_0}\,.\label{k.p_30}
\ee
This branch can have a  negative effective mass for $\gamma_1>1$, that is, a coupling $\mathcal{P}_{cv}$ between $|\mu_v\ran$ and $|c\ran$ large enough to have $2|\mathcal{P}_{cv}|^2/m_0$ larger than  the band gap $E_g$.

\subsection{Coupling to the $|\mu_c\ran$ subspace only}

To better catch the role of the orbital state degeneracy, we now consider the coupling of the 3-fold valence states $|\mu_v\ran$ to the 3-fold conduction states $|\mu_c\ran$, these two sets of states having opposite parity.

\noindent $\bullet$ Using the periodic wave functions given in Sec.~\ref{sec1.2}, we find
\be
\lan \mu'_c|\hat{p}_x|\mu_v\ran=\sum_{\vK}(K_{\mu'} G^*_{K,c})\hbar K_x\left(\frac{K_xK_yK_z}{K_\mu} G_{K,v}\right)\,,\label{k.p_31}
\ee
which differs from zero for $(\mu',\mu)$ equal to $(y,z)$ or to $(z,y)$. So, we end, since $(x,y,z)$ play the same role, with
\be
\lan y_c|\hat{p}_x|z_v\ran=\lan z_c|\hat{p}_x|y_v\ran=\hbar\sum_{\vK}K_x^2K_y^2G^*_{K,c} G_{K,v}\equiv \mathcal{Q}_{cv}\,,\label{k.p_32}
\ee
and similar results obtained from cyclic permutations. 

\begin{figure*}[t]
\begin{center}
\includegraphics[width=2\columnwidth]{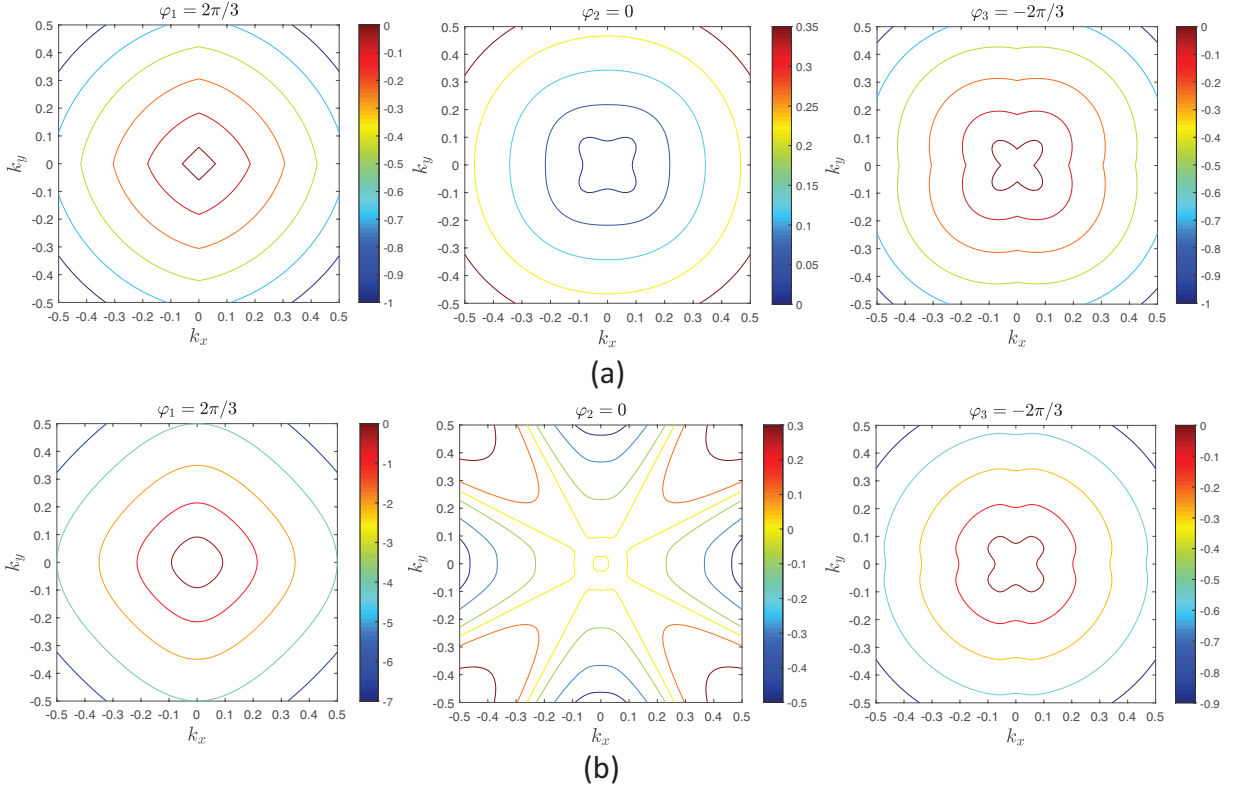}
\vspace{-2.4cm}
\caption{(a) Contours of the valence electron energies $\va_{v,\vk}-\va_{v,\textbf{0}}$ (in the unit of $\hbar^2/2m_0$) as a function of $(k_x,k_y)$ for $k_z=0.05$. They correspond to the three solutions given in Eq.~(\ref{k.p_41}) for coupling between $|\mu_v\ran$ and $|\mu_c\ran$ only. In each subfigure, the  color of each contour curve represents the energy value given by the color bar on the right. The solution for $\varphi_2=0$ corresponds to a positive  mass, as can be deduced from the fact that the $\va_{v,\vk}-\va_{v,\textbf{0}}$ energy increases from 0 with increasing $k$. The other two solutions, $\varphi_1=2\pi/3$ and $\varphi_3=-2\pi/3$, correspond to  negative masses, as seen from the fact that  the $\va_{v,\vk}-\va_{v,\textbf{0}}$ energy decreases from 0 with increasing $k$. (b) Same for  Eq.~(\ref{k.p_44}) that correspond to coupling between $|\mu_v\ran$ and both $|c\ran$ and $|\mu_c\ran$ states of the conduction band. The solution for $\varphi_2=0$ corresponds to a positive mass along $k_x= \pm k_y$ and a negative mass along $k_x=0$ or $k_y=0$; the change occurs along the yellow contour lines which correspond to $\va_{v,\vk}-\va_{v,\textbf{0}}=0$, that is, the inverse of the effective mass equal to zero. The other two solutions produce  negative masses, as $\va_{v,\vk}-\va_{v,\textbf{0}}$ decreases from 0 with increasing $k$. We have taken  $\gamma_1=16.7$ and $\gamma_3=3.5$ as for GaAs\cite{Cardona1988}.}
\label{fig:3}
\end{center}
\end{figure*}

\noindent $\bullet$ The matrix elements of the $\hat{w}_\vk$ operator between the 3-fold valence states $|\mu_v\ran$ and the 3-fold conduction states $|\mu'_c\ran$ then read as
\bea
 \lan \mu'_c|\hat{w}_\vk|x_v\ran&=&\lan \mu'_c|\frac{\hbar}{m_0}(k_x\hat{p}_x+k_y\hat{p}_y+k_z\hat{p}_z)|x_v\ran\nn\\
 &=&\frac{\hbar\mathcal{Q}_{cv}}{m_0}(k_y\delta_{\mu',z}+k_z\delta_{\mu',y})\,.\label{k.p_33}
\eea

\noindent $\bullet$ When the $P_\perp$ projector appearing in Eq.~(\ref{k.p_8}) only contains the $|\mu_c\ran$ states, we find
\bea
\mathcal{W}_{\mu',\mu}&=&\lan \mu'_v|\hat{w}_\vk\sum_{\mu''}\frac{|\mu''_c\ran\lan \mu''_c|}{-(E_g+\Delta_c)}\hat{w}_\vk|\mu_v\ran\nn\\
&\equiv& -\frac{\hbar^2}{2m_0}\gamma_3\lan \mu'_v|\hat{\mathcal{B}}^{(3)}|\mu_v\ran\,, \label{k.p_34}
\eea
where the dimensionless parameter $\gamma_3$ is given by
\be
\gamma_3=\frac{2 |\mathcal{Q}_{cv}|^2}{m_0(E_g+\Delta_c)}\,.
\ee 
The corresponding $3 \times 3$ matrix 
for $\hat{\mathcal{W}}$ then reads 
$-(\hbar^2 \gamma_3 /2m_0) \hat{\mathcal{B}}^{(3)}$ where the $\hat{\mathcal{B}}^{(3)}$ operator in the $\mu=(x,y,z)$ basis also is a symmetric real matrix 
\be
\hat{\mathcal{B}}_\mu^{(3)}=\begin{pmatrix}
k_y^2+k_z^2 & \cdot & \cdot \\
k_xk_y & k_z^2+k_x^2 & \cdot\\
k_zk_x &k_yk_z & k_x^2+k_y^2
\end{pmatrix}\,.\label{k.p_35}
\ee

The determinant of this matrix is equal to $4k_x^2k_y^2k_z^2$; so, for any $k_\mu=0$, the equation $\text{det}\hat{\mathcal{D}}=0$ has a solution $\tilde{\va}_{v,\vk}=0$, that is, valence electrons still having a positive mass $m_0$. More precisely, the solutions of $\text{det}\hat{\mathcal{D}}=0$ read as $\tilde{\va}_{v,\vk}= (\hbar^2/ 2m_0)   \gamma_3 e_3$ with $ e_3$ given by 
\bea
0&=&\left|\begin{matrix}
k_y^2+k_z^2+e_3 &  \cdot & \cdot  \\
k_xk_y & k_z^2+k_x^2+e_3 & \cdot \\
k_zk_x &k_yk_z & k_x^2+k_y^2+e_3
\end{matrix}\right|\nn\\
&=&e_3(e_3+k^2)^2+4k_x^2k_y^2k_z^2\,.\label{k.p_36}
\eea
For $k_xk_yk_z=0$, this gives
\bea
\va_{v,\vk}&=&\va_{v,\v0}+\frac{\hbar^2}{2m_0}k^2 \quad \text{non-degenerate}\,,\label{k.p_37}\\
\va_{v,\vk}&=&\va_{v,\v0}+\frac{\hbar^2}{2m_0}(1-\gamma_3)k^2 \quad \text{2-fold}\,.\label{k.p_38}
\eea
So, the coupling to $|\mu_c\ran$ only does not yet provide valence electrons with an effective mass  negative  for whatever $\vk$.

It is possible to analytically solve Eq.~(\ref{k.p_36}) for $k_xk_yk_z\not=0$ by using the Cardano's trick\cite{Nickalls}: the three solutions of $x^3+3a x-b=0$ are
\be
x^{(n)}=\sum_{\tau=\pm1} e^{i\tau \varphi_n}\left[\frac{b+\tau\sqrt{b^2+4a^3}}{2}\right]^{1/3}\label{k.p_40}
\ee
for $\varphi_n=(0,\pm2\pi/3)$. To use it, we first introduce $\Delta_3=e_3+2k^2/3$. From Eq.~(\ref{k.p_36}), we find that the resulting cubic polynomial equation  for $\Delta_3$ has no quadratic term
\be
0=\Delta_3^3-\frac{k^4}{3}\Delta_3-\frac{2k^6}{27}+4k_x^2k_y^2k_z^2\,.\label{k.p_39}
\ee
Since $k_x^2k_y^2k_z^2\leq (k^2/3)^3$, Eq.~(\ref{k.p_40})  gives the three solutions of the above equation as
\be
\Delta_3^{(n)}{=}\sum_{\tau=\pm1}e^{i\tau \varphi_n}\!\Bigg[\!\left(\frac{k^2}{3}\right)^3\!\!{-}2P_\vk+2i\tau  \sqrt{P_\vk\left(\frac{k^2}{3}\right)^3\!\!{-}P^2_\vk} \Bigg]^{1/3},\label{k.p_41}
\ee
with $P_\vk=k_x^2k_y^2k_z^2$. The shapes of the resulting energy contours for $\va_{v,\vk}$ are shown in Fig.~\ref{fig:3}(a).

\subsection{Coupling to both $|c\ran$ and $|\mu_c\ran$ states}
 
 As now shown, the  coupling to the $|c\ran$ and $|\mu_c\ran$ states not only decreases the number of  $\vk$ momenta at which the valence electron effective mass stays equal to $m_0$, but this coupling  also generates a warping dependence of the dispersion relation in $S_\vk=k_x^2k_y^2+k_y^2k_z^2+k_z^2k_x^2$.

 \noindent $\bullet$ For $P_\perp=|c\ran\lan c|+\sum_\mu |\mu_c\ran\lan \mu_c|$, we find
 \be
\mathcal{W}_{\mu',\mu}=-\frac{\hbar^2}{2m_0}\left(\gamma_1 \mathcal{B}^{(1)}_{\mu',\mu}+\gamma_3  \mathcal{B}^{(3)}_{\mu',\mu}\right)\equiv -\frac{\hbar^2}{2m_0} \lan\mu'_v| \hat{\mathcal{B}}|\mu_v\ran\,.\label{k.p_42}
\ee
Using the $ \hat{\mathcal{B}}_\mu^{(1)}$ and $ \hat{\mathcal{B}}_\mu^{(3)}$ matrices calculated above, we can write $\hat{\mathcal{B}}_\mu$ as $\gamma_3k^2 \mathcal{I}+ \hat{\mathcal{C}}_\mu$ where $\mathcal{I}$ is the $3\times3$ unit matrix while the matrix $\hat{\mathcal{C}}_\mu$ is given by
\be
\hat{\mathcal{C}}_\mu=\begin{pmatrix}
k_x^2 \gamma_-& \cdot & \cdot \\
k_xk_y \gamma_+& k_y^2 \gamma_-& \cdot\\
k_zk_x \gamma_+&k_yk_z \gamma_+&k_z^2\gamma_-
\end{pmatrix}\,,\label{k.p_43}
\ee
with $\gamma_{\pm}=\gamma_1\pm\gamma_3$.

 \noindent $\bullet$ By writing $\tilde{\va}_{v,\vk}$ as $(\hbar^2/2m_0)e$ with $e=e'-\gamma_3k^2$, we find that the valence electron eigenvalues resulting from the couplings to the $|c\ran$ and $|\mu_c\ran$ conduction states follow from
 \be
 0= e'^3+(\gamma_1-\gamma_3)k^2 e'^2-4\gamma_1\gamma_3S_\vk e'+4\gamma_3^2(3\gamma_1+\gamma_3)P_\vk\,,
 \label{k.p_44}
 \ee
 with $(S_\vk,P_\vk)$ defined in Eq.~(\ref{k.p_1}). The above equation demonstrates a warping behavior through the $S_\vk$ term; note that both the $\gamma_1$ and $\gamma_3$ couplings to the $|c\ran$ and $|\mu_c\ran$ conduction states are necessary to bring the warping into the problem. Moreover, we see that when these two couplings exist, that is, when $\gamma_1\gamma_3\not=0$, the above equation has a solution $e'=\gamma_3k^2$, that is, $e=0$, for
\be
k_x=0, \qquad k_y=\pm k_z\,,\label{k.p_45}
\ee
and their cyclic permutations. So, for some $k$ values, the valence electron effective mass still is positive and equal to $m_0$. 

For arbitrary $k$'s, we can obtain the three solutions of Eq.~(\ref{k.p_44}) by again using the Cardano's trick. The shapes of the energy contours for $\va_{v,\vk}$ are shown in Fig.~\ref{fig:3}(b).\

All this shows that including the couplings to the $|c\ran$ state and the $|\mu_c\ran$ states of the conduction band reduces the number of $\vk$ momenta at which the valence electron effective mass stays equal to $m_0$. However, this mixing of orbital symmetries is not enough to produce a negative curvature for whatever $\vk$. We are going to show that an additional mixing, that comes from the spin-orbit interaction, is necessary to endow a valence electron with a negative effective mass whatever its $\vk$ momentum.

\subsection{$\hat{\mathcal{B}}^{(1)}$ and $\hat{\mathcal{B}}^{(3)}$ operators in the $(1,0,-1)$ basis}
To derive the effective masses of valence electrons in the presence of spin, we have to turn from the $(x,y,z)$ crystal basis, for which calculations are easy to perform with the help of cyclic permutations, to the $(1,0,-1)$ basis for which such cyclic symmetry is broken by the spin quantization axis.

 \noindent $\bullet$  By inserting the $\hat{\mathcal{B}}_\mu^{(1)}$ matrix elements in the $(x,y,z)$ basis, given in Eq.~(\ref{k.p_28}), into the general link given in Eq.~(\ref{k.p_22}) between matrices in different bases, we obtain the $\hat{\mathcal{B}}^{(1)}$ operator in the $(1,0,-1)$ basis as 
\be
\hat{\mathcal{B}}_\eta^{(1)}=\begin{pmatrix}
|k_1|^2 & \cdot & \cdot \\
k_0^*k_1 & |k_0|^2 & \cdot\\
k^*_{-1}k_1 &k_{-1}^*k_0 & |k_{-1}|^2
\end{pmatrix}\,,\label{k.p_46}
\ee
with $k_{\pm1}=(\mp i k_x+k_y)/\sqrt{2}$ and $k_0=ik_z$. We can check that $\text{det}\hat{\mathcal{B}}_\eta^{(1)}=0$ as expected, since the determinant is invariant under a basis change. We can also recover the eigenvalue equation for $e_1$ given in Eq.~(\ref{k.p_29}) but in a far heavier way, due to the broken symmetry of the $(1,0,-1)$ basis along $z$.\

 \noindent $\bullet$ In the same way, changing from $(x,y,z)$ to $(1,0,-1)$ basis gives the $\hat{\mathcal{B}}^{(3)}$ operator in the $(1,0,-1)$ basis as
 \be
 \hat{\mathcal{B}}_\eta^{(3)}=\begin{pmatrix}
|k_1|^2 +|k_0|^2 & \cdot & \cdot \\
k_0^*k_1 & 2|k_1|^2 & \cdot\\
-{k^*_1}^2 &k_0k_1 & |k_{-1}|^2+|k_0|^2
\end{pmatrix}\,.\label{k.p_47}
 \ee
It still is rather easy to check that the determinant of this matrix is equal to $4P_\vk$, but the derivation of Eq.~(\ref{k.p_36}) for $e_3$ takes more time.

It can be of interest to note that the off-diagonal terms of the $\hat{\mathcal{B}}^{(1)}$ and $\hat{\mathcal{B}}^{(3)}$ operators are identical when written in the $(x,y,z)$ basis, while only two of them are identical when written in the $(1,0,-1)$ basis. Difference between $\mathcal{B}^{(1)}_{-1,1}$ and $\mathcal{B}^{(3)}_{-1,1}$ comes from the broken symmetry along the spin quantization axis which produces different diagonal terms, $(-\mathcal{B}^{(1)}_{x,x}+\mathcal{B}^{(1)}_{y,y})/2$ and $(-\mathcal{B}^{(3)}_{x,x}+\mathcal{B}^{(3)}_{y,y})/2$, in the $(1,-1)$ couplings (see Eq.~(\ref{k.p_22})).

\section{ Spin-orbit effect on valence effective masses \label{sec5}}

\subsection{From the $|\eta\ran\otimes |\pm1/2\ran$ to the $|\tilde{\jmath}\ran$ states\label{5.1}}

The first step to handle the effect of spin-orbit coupling on the valence states is to go from the $|\eta\ran\otimes |\pm1/2\ran$ states to their linear combinations $|\tilde{\jmath}\ran$ which are eigenstate of the spin-orbit interaction. In the following, we consider that the spin-orbit splitting $\Delta_{so}$ is large enough and the $\vk$ momentum small enough, to possibly restrict the spin-orbit eigenstates to the 4-fold states $|\tilde{\jmath}=3/2\ran$ given in Eqs.~(\ref{k.p_13},\ref{k.p_14}), in spite of the fact that these states are coupled to the $|\tilde{\jmath}=1/2\ran$ states which have the same parity, through their common coupling to the conduction states $|c\ran$ or $|\mu_c\ran$.

For $\hat{A}$ operator that does not act on spin, like the $\hat{w}_\vk$ operator, the parts of the $|\tilde{\jmath}\ran$ states that are coupled must have the same spin. Accordingly, we get, using the $\hat{A}_\eta$ matrix elements defined in Eq.~(\ref{k.p_19}), 
\be
A_{\frac{3}{2},\frac{3}{2}}=\left\lan 3/2\right|\hat{A} \left|3/2\right\ran=\left\lan 1/2\right|\otimes\left\lan 1 \right| \hat{A}\left|1\right\ran\otimes \left|1/2\right\ran=A_{1,1}\,.\label{k.p_48}
\ee
In the same way
\bea
A_{\frac{1}{2},\frac{3}{2}}&=&\frac{\left\lan -1/2\right|\otimes\left\lan 1 \right|+\sqrt{2}\left\lan 1/2\right|\otimes\left\lan 0 \right| }{\sqrt{3}}\hat{A}\left|1\right\ran\otimes \left|1/2\right\ran\nn\\
&=&\sqrt{\frac{2}{3}}A_{0,1}\,,\label{k.p_49}
\eea
and so on... 

The $\hat{A}$ operator in the $(3/2,1/2,-1/2,-3/2)$ basis then appears as 
\be
\hat{A}_{\tilde{\jmath}}=\begin{pmatrix}
A_{1,1} & \cdot & \cdot  & \cdot \\
 \sqrt{\frac{2}{3}}A_{0,1} & \frac{A_{1,1}+2A_{0,0}}{3} & \cdot  & \cdot  \\
\frac{1}{\sqrt{3}}A_{-1,1} &\sqrt{2}\frac{A_{0,1}+A_{-1,0}}{3}  & \frac{A_{1,1}+2A_{0,0}}{3}&\cdot \\
0 & \frac{1}{\sqrt{3}}A_{-1,1} &   \sqrt{\frac{2}{3}}A_{-1,0} & A_{-1,-1}
\end{pmatrix}\,.\label{k.p_50}
\ee
Note that for real $A_{\mu',\mu}$'s, as in the case of present interest,
\be
A_{0,1}+A_{-1,0}=\frac{-A_{z,x}-iA_{z,y}}{\sqrt{2}}+\frac{A_{x,z}+iA_{y,z}}{\sqrt{2}}=0\,.\label{k.p_51}
\ee
So, all elements in the second diagonal are equal to zero.

\subsection{Coupling to $|c\ran\otimes |\pm1/2\ran$ states only}

From the $\hat{\mathcal{B}}^{(1)}_{\eta}$
 matrix in the $3\times3$ orbital subspace $(1,0,-1)$, given in Eq.~(\ref{k.p_46}), we can derive its expression in the $4\times4$ subspace $(3/2,1/2,-1/2,-3/2)$, by using Eq.~(\ref{k.p_50}). We find
\be
\hat{\mathcal{B}}^{(1)}_{\tilde{\jmath}}
=\begin{pmatrix}
|k_1|^2 & \cdot & \cdot & \cdot \\
 \sqrt{\frac{2}{3}}k_0^*k_1 & \frac{|k_1|^2{+}2|k_0|^2}{3} & \cdot & \cdot \\
\frac{1}{\sqrt{3}}k_1^2 &0  & \frac{|k_1|^2{+}2|k_0|^2}{3}& \cdot\\
0 & \frac{1}{\sqrt{3}}k_1^2 &   \sqrt{\frac{2}{3}}k_0k_1 &|k_1|^2 
\end{pmatrix}\,.\label{k.p_52}
\ee
The determinant of this matrix is equal to zero; so, there still is a solution with a positive valence electron mass  $m_0$. More precisely, the eigenvalue equation obtained from this matrix reads
\be
0=e_1^2 (e_1^2+2k^2/3)^2\,.\label{k.p_53}
\ee
It has a 2-fold solution $e_1=0$, that leads to $\va_{v,\vk}=\va_{v,{\bf0}}+\hbar^2 k^2/2m_0$ and a 2-fold solution $e_1=-2k^2/3$, that leads to  
\be
\va_{v,\vk}=\va_{v,{\bf0}}+\frac{\hbar^2 k^2}{2m_0}(1- \frac{2}{3} \gamma_1)\,.\label{k.p_54}
\ee

 This shows that the coupling to the non-degenerate conduction level splits the upper 4-fold valence states resulting from spin-orbit interaction, into two states still having a positive effective mass equal to $m_0$ and two states with an effective mass possibly negative for a coupling $\gamma_1$ large enough. The effect of the spin-orbit interaction slightly reduces this possibility through a prefactor increase from $(1-\gamma_1)$ to $(1-2\gamma_1/3)$.
 
 \begin{figure}[t]
\begin{center}
\subfigure[]{\label{fig:S2a} \includegraphics[trim=0cm 0cm 0cm 0cm,clip,width=2.9in]{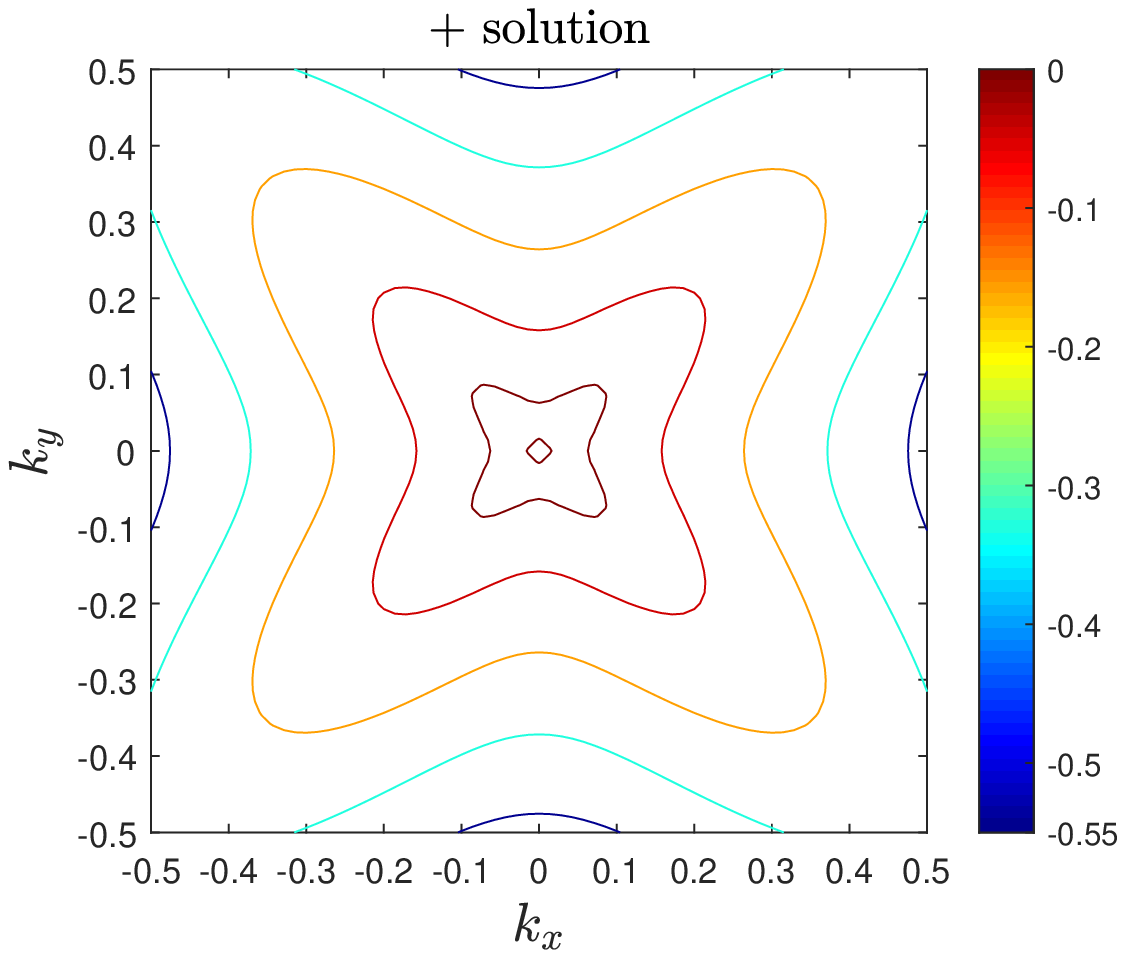}}
\subfigure[]{\label{fig:S2b} \includegraphics[trim=0cm 0cm 0cm 0cm,clip,width=2.9in]{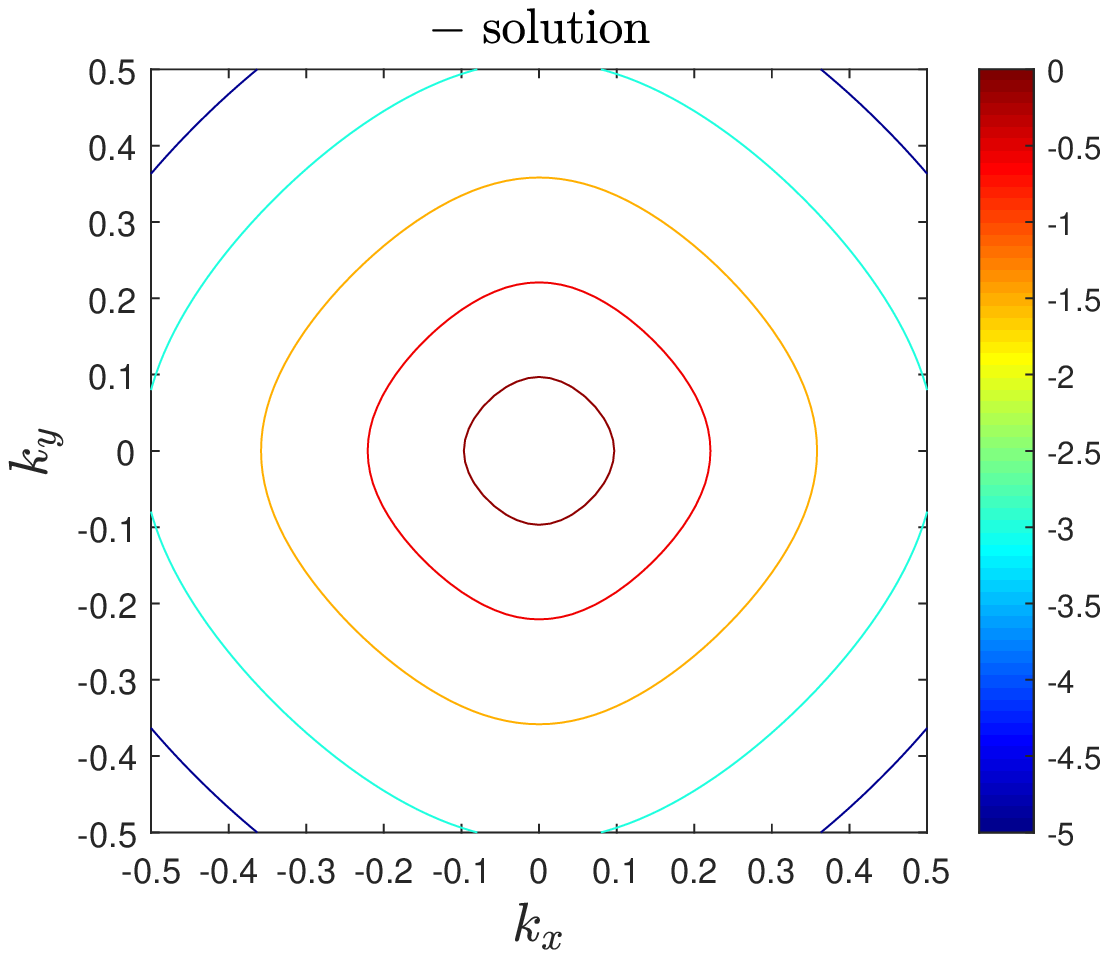}}
\end{center}
\vspace{-0.5cm}
\caption{\small Energy contours of $\va_{v,\vk}-\va_{v,\textbf{0}}$ (in the unit of $\hbar^2/2m_0$) corresponding to (a) the $+$ solution and (b) the $-$ solution given in Eq.~(\ref{k.p_60}). Parameters are the same as in Fig.~\ref{fig:3} for GaAs.  The two solutions correspond to a decrease of $\va_{v,\vk}-\va_{v,\textbf{0}}$ from 0 for whatever $\vk$, that is, a negative effective mass for all $\vk$ directions.}
\label{fig:4}
\end{figure}

\subsection{Coupling to $|\mu_c\ran\otimes |\pm1/2\ran$ only }

For a spin-orbit splitting small compared to the band gap, we can neglect its effect on the 3-fold orbital states $|\mu_c\ran$ of the conduction band, and consider that the six states  $|\mu_c\ran\otimes |\pm1/2\ran$ have the same energy $(E_g+\Delta_c)$.

From the $\hat{\mathcal{B}}^{(3)}_{\eta}$  matrix in the $3\times3$ orbital subspace $(1,0,-1)$, given in Eq.~(\ref{k.p_47}), we can derive its expression in the $4\times4$ subspace $(3/2,1/2,-1/2,-3/2)$, by using Eq.~(\ref{k.p_50}). We find
\be
\hat{\mathcal{B}}^{(3)}_{\tilde{\jmath}}
{=}\!\begin{pmatrix}
|k_1|^2{+}|k_0|^2 & \cdot & \cdot & \cdot \\
 \sqrt{\frac{2}{3}}k_0^*k_1 & \frac{5|k_1|^2{+}|k_0|^2}{3} & \cdot & \cdot \\
-\frac{1}{\sqrt{3}}{k_1^*}^2 &0  & \frac{5|k_1|^2{+}|k_0|^2}{3}& \cdot\\
0 & -\frac{1}{\sqrt{3}}{k_1^*}^2 &   \sqrt{\frac{2}{3}}k_0k_1 &|k_1|^2 {+}|k_0|^2 
\end{pmatrix}.\label{k.p_55}
\ee
This matrix has obvious similarities with $\hat{\mathcal{B}}^{(1)}_{\tilde{\jmath}}$, but some crucial differences. These differences bring to $\hat{\mathcal{B}}^{(3)}_{\tilde{\jmath}}$ a determinant $k^8/9$ which differs from zero for \textit{all} directions of the electron $\vk$ momentum. As a result, there is no more solution  $\tilde{\va}_{v,\vk}=0$: all the curvatures in the valence dispersion relation now differ from the one of a free electron.

By writing  $\tilde{\va}_{v,\vk}$ as $ (\hbar^2/ 2m_0)\gamma_3e_3$, we find, from the determinant of the above matrix, the equation for $e_3$ as  
\be
0=\left(e_3^2+4k^2e_3/3+k^4/3\right)^2=\left(e_3+k^2\right)^2\left(e_3+k^2/3\right)^2\,.\label{k.p_56}
\ee
The resulting energies for the fourfold $|\tilde{\jmath}=3/2\ran$ states then are 
\bea
\va_{v,\vk}&=&\va_{v,\v0}+\frac{\hbar^2}{2m_0}(1-\gamma_3)k^2 \quad \text{2-fold}\,,\label{k.p_57}\\
\va_{v,\vk}&=&\va_{v,\v0}+\frac{\hbar^2}{2m_0}(1-\frac{\gamma_3}{3})k^2 \quad \text{2-fold}\,.\label{k.p_58}
\eea
These valence electrons all have a negative effective mass for a large enough  coupling $\gamma_3$ between the $|\mu_v\ran$ and $|\mu_c\ran$ states.

This  shows that in order to have a negative valence mass whatever the direction of the $\vk$ momentum, it is necessary to mix the $|\mu_v\ran$ valence orbital states not only through their coupling to the 3-fold orbital states $|\mu_c\ran$ of the conduction band, but also between themselves through the spin states via the spin-orbit interaction.

\subsection{Coupling to $|c\ran\otimes |\pm1/2\ran$ and $|\mu_c\ran\otimes |\pm1/2\ran$ states\label{sec5d}}

As in the absence of spin, we write  the energy change as $\tilde{\va}_{v,\vk}=( \hbar^2/ 2m_0  )e$ with $e=(e'-\gamma_3k^2)$. We find that $e'$ follows from the cancellation of the $4\times4$ determinant given by
\begin{widetext}
\bea
0&=&\left|\begin{matrix}
|k_1|^2\gamma_{-}+e' &   \cdot  & \cdot & \cdot  \\
\sqrt{\frac{2}{3}}k_0^* k_1 \gamma_{+}& \frac{|k_1|^2+2|k_0|^2}{3}\gamma_{-}+e' & \cdot & \cdot  \\
\frac{\gamma_1k_1^2-\gamma_3{k_1^*}^2}{\sqrt{3}} &0 & \frac{|k_1|^2+2|k_0|^2}{3}\gamma_{-}+e' & \cdot\\
0 & \frac{\gamma_1k_1^2-\gamma_3{k_1^*}^2}{\sqrt{3}} &  \sqrt{\frac{2}{3}}k_0 k_1  \gamma_{+}&|k_1|^2 \gamma_{-}+e'
\end{matrix}\right|\,.\label{k.p_59}
\eea
\end{widetext}
The resulting equation reads
\be
0=\left( e'^2+ \frac{2}{3}(\gamma_1-\gamma_3)k^2e'- \frac{4}{3}\gamma_1\gamma_3S_\vk\right)^2.
\ee
From its solutions, we find that the 4-fold states $|\tilde{\jmath}=3/2\ran$ split as two sets of 2-fold states with energies 
\bea
\va_{v,\vk}=\va_{v,\v0}+\frac{\hbar^2}{2m_0}\Big[(1-\frac{\gamma_1+2\gamma_3}{3})k^2\label{k.p_60}\,\,\,\,\,\,\,\,\,\,\,\,\,\,\,\,\,\,\,\,\,\,\,\,\,\,\,\,\,\,\,\,\,\,\,\,\,\,\,\,
\\
\pm\frac{1}{3}\sqrt{(\gamma_1-\gamma_3)^2k^4+12 \gamma_1\gamma_3S_\vk}\Big] \,.
\nn
\eea
These results were first derived by Dresselhaus, Kip and Kittel \cite{DresselhausPR1955}. The shapes of the energy contours for  $\va_{v,\vk}$  are shown in Fig.~\ref{fig:4}.

 It is easy to check that the above equation leads to the energies obtained in Eq.~(\ref{k.p_54}) when $\gamma_3=0$ and the energies obtained in Eqs.~(\ref{k.p_57},\ref{k.p_58}) when $\gamma_1=0$. These results also confirm our previous observation that in order to have a negative valence electron mass whatever $\vk$, it is necessary to   mix the $|\mu_c\ran$ valence states with the spin states through the spin-orbit interaction, and to couple  these valence states to the 3-fold  conduction states $|\mu_c\ran$ through the $\gamma_3$ parameter.

\section{Conclusion}
It is commonly accepted that a periodic ion potential brings to semiconductor electrons a mass that  drastically turns negative close to a band maximum. Yet, the precise physics for why electrons in a 3-fold valence orbital state all acquire a negative effective mass for whatever their momentum direction, is quite tricky. In this work, we  consider  various scenarios of couplings between 3-fold valence orbital states and conduction orbital states with different degeneracies that are close to the band gap, as well as coupling to spin states through spin-orbit interaction. Our analysis, based on a $\vk\cdot\vp$ approach, proves that while the coupling to  conduction levels is the primary reason for driving the mass of the valence electrons toward a negative value, a further symmetry breaking of the  valence orbital degeneracy along the spin quantization axis as a result of the spin-orbit interaction, is necessary to ensure that \textit{all} valence electrons acquire a negative mass \textit{whatever} the $\vk$ momentum direction. Since  spin-orbit interaction is a relativistic effect, the negative effective mass of  semiconductor valence electrons thus constitutes an unexpected signature of quantum relativity, unrevealed until now.

\end{document}